# Understanding User Behavior in Volumetric Video Watching: Dataset, Analysis and Prediction

Kaiyuan Hu[1,2†], Haowen Yang[1,6†], Yili Jin[5], Junhua Liu[1]
Yongting Chen[1,2], Miao Zhang[4], Fangxin Wang[2,1,3*]

[1]The Future Network of Intelligence Institute, The Chinese University of Hong Kong, Shenzhen
[2]School of Science and Engineering, The Chinese University of Hong Kong, Shenzhen
[3]The Guangdong Provincial Key Laboratory of Future Networks of Intelligence
[4]Simon Fraser University, [5]McGill University, [6]Versee Inc.

## ABSTRACT

Volumetric video emerges as a new attractive video paradigm in recent years since it provides an immersive and interactive 3D viewing experience with six degree-of-freedom (DoF). Unlike traditional 2D or panoramic videos, volumetric videos require dense point clouds, voxels, meshes, or huge neural models to depict volumetric scenes, which results in a prohibitively high bandwidth burden for video delivery. Users' behavior analysis, especially the viewport and gaze analysis, then plays a significant role in prioritizing the content streaming within users' viewport and degrading the remaining content to maximize user QoE with limited bandwidth. Although understanding user behavior is crucial, to the best of our best knowledge, there are no available 3D volumetric video viewing datasets containing fine-grained user interactivity features, not to mention further analysis and behavior prediction.

In this paper, we for the first time release a volumetric video viewing behavior dataset, with a large scale, multiple dimensions, and diverse conditions. We conduct an in-depth analysis to understand user behaviors when viewing volumetric videos. Interesting findings on user viewport, gaze, and motion preference related to different videos and users are revealed. We finally design a transformer-based viewport prediction model that fuses the features of both gaze and motion, which is able to achieve high accuracy at various conditions. Our prediction model is expected to further benefit volumetric video streaming optimization.

Our dataset, along with the corresponding visualization tools is accessible at https://cuhksz-inml.github.io/user-behavior-in-vv-watching/

## CCS CONCEPTS

• **Human-centered computing** → **Virtual reality**; • **Information systems** → **Multimedia databases**.

*Fangxin Wang is the corresponding author.
†Both authors contributed equally to this research.

## KEYWORDS

Volumetric videos, Dataset, User Behavior Analysis

## 1 INTRODUCTION

The confluence of video and the recently booming 3D representation technology embraces a new video paradigm, i.e., the *volumetric video* (VV). Different from traditional 2D video that has mature codecs based on frames and pixels, volumetric video is still in its infant stage with various representation formats, such as point cloud [11, 34], voxel [33], mesh [34], and even neural representations [23]. Volumetric video is envisioned as a fundamental service that is able to facilitate various new applications such as extended reality (XR) and Metaverse, empowering entertainment [20], healthcare [7], and education [2], etc. The global industry VV market is expected to reach 22.5 billion USD by 2024 [21].

Unlike traditional or 360-degree videos that only provide flat or curved 2D experience, volumetric video captures the scene and objects in 3D format, providing 6 degree-of-freedom (DoF) viewing experience, including three dimensions of position (X, Y, Z) and three dimensions of orientation (yaw, pitch, roll). This new viewing paradigm revolutionizes the way we consume video content, offering an unprecedented full immersive and interactive experience. Such interactivity between the user and the 3D video already demonstrates great value in various fields, e.g., revealing mental activity, inferring user preference, and even identifying different users.

Due to the extreme complexity in volumetric video representation, e.g., extensive points or meshes using point cloud or 3D mesh formats, or huge neural models using implicit neural representation, the size of a volumetric video is usually much larger (up to 100x) than the 2D representation in the same condition. Thus, streaming volumetric video through the current network infrastructure tends to become a key challenge. Users' behavior analysis, especially the field of view (FoV) and gaze analysis, then plays a significant role because we can prioritize the content streaming within FoV and reduce or even ignore the content out of FoV to maximize user's QoE with limited network transmission capacity [8].

Table 1: User Information

| Gender | Female | | Male | |
|---|---|---|---|---|
| | 27 | | 23 | |
| Age | 16-20 | 20-24 | 24-30 | 30+ |
| | 25 | 17 | 5 | 3 |
| VR Exp (Times) | Never | 1-5 | 6-10 | 10+ |
| | 32 | 9 | 6 | 3 |
| VV Exp (Times) | Never | 1-5 | 6-10 | 10+ |
| | 41 | 3 | 3 | 3 |

Although understanding user behavior is crucial, to our best knowledge, there is no available 3D volumetric video viewing datasets containing fine-grained user interactivity features. Pioneer researchers in the community of multimedia have contributed some 3D datasets on objects or scenes [9, 26], but they never focus on the analysis and understanding of user behavior in volumetric video. Thus, an open dataset in this context is in urgent need to reveal the viewing characteristics, optimize the video streaming, and further facilitate the research in the related community.

In this paper, we propose the first large-scale user behavior dataset on volumetric video viewing with rich dimensions across various scenes, including the six DoF viewport, gaze, and motion features. We next conduct a comprehensive data analysis to deeply understand the user behavior, fully capture the potential correlations among viewport, gaze, and motion trajectory, and further reveal the future viewing activity. We find that VV users exhibit distinct regions of interest and display varying movement patterns based on different scenarios and personalities. Based on our observations and findings, we conduct a pilot study on viewport adaptive 3D volumetric video streaming. We design a transformer-based model to well capture the inherent relationship between the motion and gaze, and further achieve an accurate and robust viewport prediction for video streaming optimization.

The contributions of our work are summarized as follows:
▷ We for the first time release a volumetric video viewing behavior dataset, with large scale (50 users), multiple dimensions (8 attributes), and diverse conditions (including both static and dynamic scenes, both single and multi-user activities).
▷ We conduct an in-depth analysis to understand user behaviors when viewing volumetric videos. Interesting findings on user viewport, gaze, and motion preference related to different videos and users are revealed.
▷ We design a transformer-based viewport prediction model that fuses the features of both gaze and motion, which is able to achieve high accuracy and strong robustness.

The rest of this paper is organized as follows. Section 2 gives an overall description of the dataset, including how data is collected as well as the video and dataset attribute description. Section 3 gives an initial visualization of the dataset, plotting headset movement and gaze direction. Section 4 introduces our analysis of user behavior in detail, and also reveals some interesting findings based on our observation. Motivated by these, section 5 proposes a transformer-based viewport prediction for six DoF volumetric video viewing. We further give some potential applications in section 6 and conclude this work in section 7.

## 2 DATASET

In this section, we introduce the details of our dataset regarding the collection procedure, dataset description, and user information.

### 2.1 Data Collection Procedure

For convenience, we select volumetric videos from the current most appropriate public volumetric dataset FSVVD [9] related to our context, which contains 26 volumetric videos represented by point cloud covering multiple common scenarios such as education, exercise, daily life, and entertainment. We seek 50 volunteers to participate in this dataset collection. These volunteers are given enough time and guidance to get familiar with the 3D volumetric environment. Videos are preloaded and played through Unity[1] when a volunteer is wearing a Meta Quest Pro[2] headset. People are able to freely navigate the 3D scenes and watch the activities from any viewing angle and any position within a 5x5 square meters space, as required by the FSVVD video dataset.

The VR headset has a built-in accelerometer and we are able to easily calculate the current headset position (X,Y,Z) and the rotation of the headset (yaw, pitch, and roll). Besides, gaze information is also important as it provides more fine-grained features [12]. For the gaze data collection, we rely on the built-in eye tracker in the headset with a sample rate of 144 Hz. The collected data consisted of 8 dimensions, including 3 rotational angles corresponding to the position of each eye, plus the confidence level. Since there are subtle differences (usually less than 3°) in the gaze data between the two eyes, we use the weighted average of the two eyes as the gaze in our later analysis.

### 2.2 Viewer Selection

Different viewers can also have quite personalized preferences on the same video content and conduct diverse behaviors. Therefore, we try our best to choose volunteers with different backgrounds, majors, hobbies, ages, genders, and familiarity levels with VR. Detailed information is listed in Table 1. Once the recording ends, the volunteers are asked to fill out a questionnaire about these information, and the overall experience of watching volumetric videos.

### 2.3 Video Selection

We argue that the video content should have a significant impact on the viewer's behavior feature. A viewer's attention can largely change if provided with different video content. To analyze the impact of video content on users, we selected 6 different scenes aiming to cover more representative scenarios. Specifically, we mainly evaluate the impact of actor numbers and the movement level of the actors. We divide the movement of target actors as spatial movement (e.g., moving from one position to another) and self-movement (e.g., body movement without obvious position change). Table. 2 indicates the detailed taxonomy of our selected video.

---
[1]https://unity.com/
[2]https://www.meta.com/quest

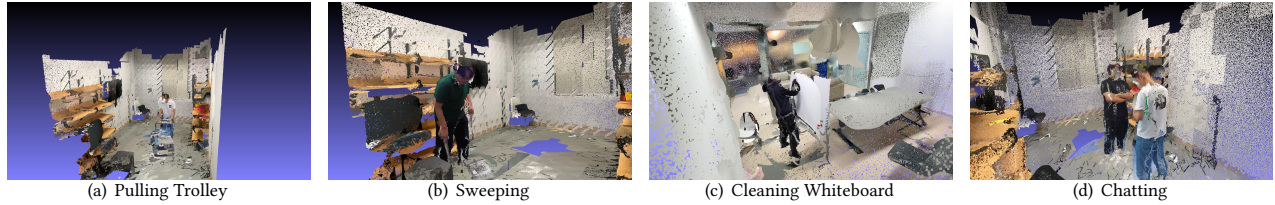

Figure 1: Example of used VV: 'Pulling Trolley', 'Sweeping', 'Cleaning Whiteboard', and 'Chatting'

Table 2: Description of selected volumetric videos:

| Name | #Actors | Spatial Movements | Body Movements | Environment Interaction | #Frame |
|---|---|---|---|---|---|
| Chatting | 2 | Small | Small | - | 300 |
| Cleaning Whiteboard | 1 | Static | Large | ✓ | 300 |
| News Interviewing | 2 | Small | Small | - | 300 |
| Pulling Trolley | 1 | Large | Small | ✓ | 300 |
| Presenting | 2 | Static | Small | - | 300 |
| Sweeping | 1 | Middle | Middle | ✓ | 300 |

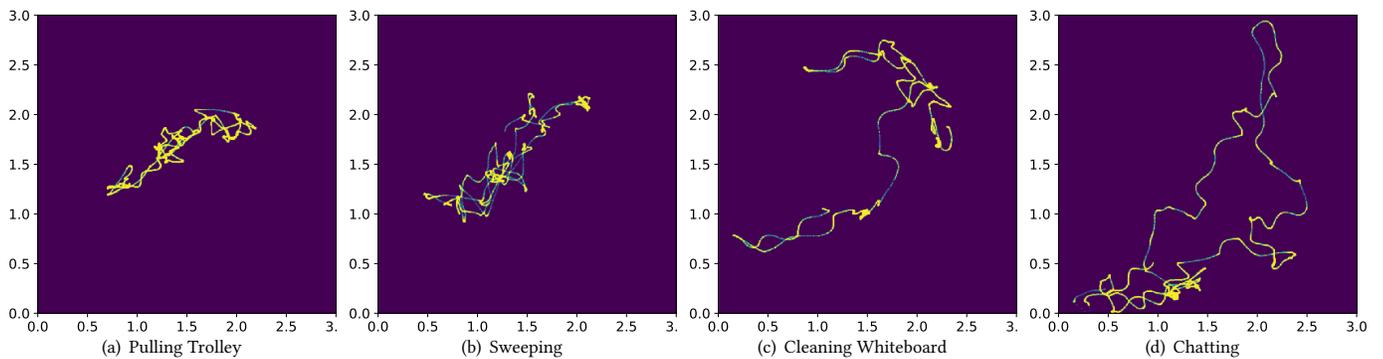

Figure 2: The aerial view for movement trajectory heatmap of different volumetric scenes. The lighter yellow color indicates a longer dwelling time and vice versa for the darker blue color.

## 2.4 Dataset Description

Our collected dataset consists of 28 dimensions, including the frame number and time stamp of each sample, the spatial movement (the spatial coordinates of X, Y, and Z axes), the rotational orientation (rotation angles of Yaw, Pitch, and Roll) information of the headset and two wireless controllers, and the gaze information of both eyes with two confidence indexes.

## 3 VISUALIZATION

To help better understand our dataset and promote further study, we first give a visualization of the dataset and provide preliminary analysis on headset movement and gaze information. We select four representative scenes for observation and subsequent analysis, i.e., pulling trolley, sweeping, cleaning whiteboard, and chatting.

### 3.1 Headset Movement Trajectory

We first observe the user movement, represented by the headset movement trajectory in our dataset. Among all the participating volunteers, we randomly select one and compare his/her movement trajectory. According to our observation, the values at the Z axis almost keep stable. This is because people rarely crouch down and stand up, which follows our intuition about people's behaviors.

Thus, we select to use an aerial view to better depict the trajectory. Fig. 2(a) shows the heatmap of movement trajectory across different scenes from a randomly selected user. Some interesting findings can be obtained. For 'Pulling Trolley' in Fig. 2(a) and 'Sweeping' in Fig. 2(b), the movement trajectories are relatively uniform and concentrated, indicating a slow movement within a small region. This matches our findings that **for volumetric videos with large movement, viewers tend to follow the moving object and are prone to pay more attention therein**. While for 'Cleaning Whiteboard' in Fig. 2(c) and 'Chatting' in Fig. 2(d), the trajectory is more dispersive. This indicates that **for small-movement or even static scenes, viewers may go around and observe the object more from different angles**.

### 3.2 Gaze Direction

Users' gaze information is also a significant indicator of user VV interactivity. We then try to visualize the gaze direction in our dataset. However, different from the traditional 2D video, the gaze can be simply projected onto the video surface, in 3D volumetric scene, the starting point of the gaze is changing along with the movement. Therefore, we need to combine these two together.

Since the rotational angles returned from the headset are represented using degrees in Euler angles, for the convenience of subsequent calculation and visualization, we transform the data into a rotation matrix. We convert the angle into radians to compute the viewport area, where $\alpha$, $\beta$, and $\gamma$ stand for yaw, pitch, and roll, respectively.

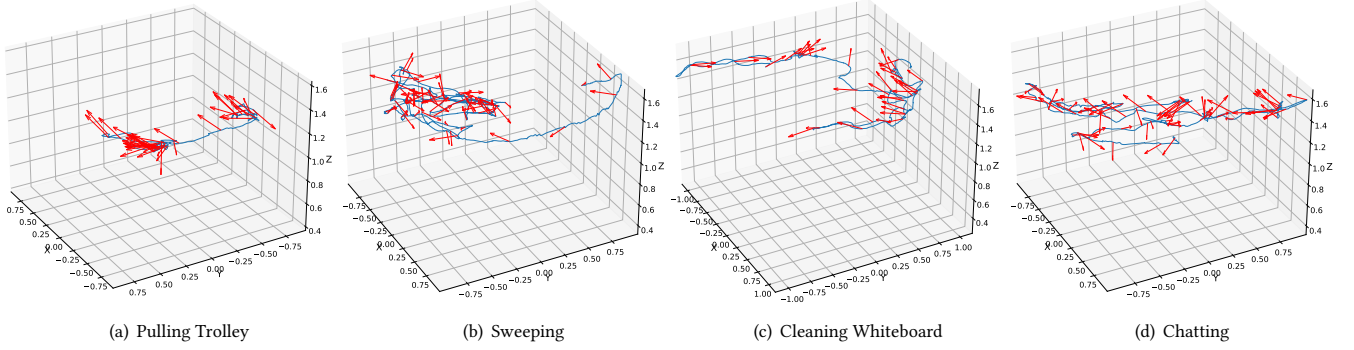

(a) Pulling Trolley	(b) Sweeping	(c) Cleaning Whiteboard	(d) Chatting

**Figure 3: Gaze Direction with Movement Trajectory. The blue line represents the movement trajectory and the red arrows indicate the gaze direction.**

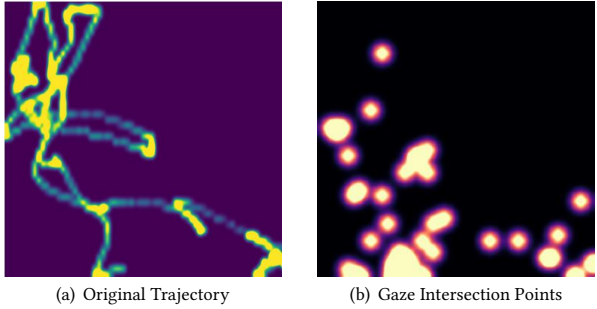

(a) Original Trajectory	(b) Gaze Intersection Points

**Figure 4: Illustration of the intersection between movement trajectory and gaze ray. The left figure shows the original user movement trajectory, and the right figure indicates the point where the movement trajectory coincides with the gaze ray.**

As denoted in Eq. 1, matrix $R$ comprises the product of the rotation matrices about the yaw, pitch, and roll axes to represent the rotation matrix of users' headset movement and gaze movement.

$$R = \begin{bmatrix} 1 & 0 & 0 \\ 0 & \cos\gamma & -\sin\gamma \\ 0 & \sin\gamma & \cos\gamma \end{bmatrix} \begin{bmatrix} \cos\beta & 0 & \sin\beta \\ 0 & 1 & 0 \\ -\sin\beta & 0 & \cos\beta \end{bmatrix} \begin{bmatrix} \cos\alpha & \sin\alpha & 0 \\ \sin\alpha & \cos\alpha & 0 \\ 0 & 0 & 1 \end{bmatrix} \quad (1)$$

Using the above transformation formula, we get the rotation matrix of both the gaze and the headset. To transform the gaze direction from the local coordinate system of the headset to the global coordinate system, we apply a rotation matrix using the orientation data provided by the headset, representing the transformation from the local coordinate system of the headset to the global coordinate system. The above transformation could be represented as:

$$R_g = R_h * R_e \quad (2)$$

where $R_g$ represents the global gaze orientation matrix, $R_h$ and $R_e$ represent the headset and local gaze (eye) orientation matrix respectively.

The combination of gaze direction and the headsets' movement trajectory is visualized in Fig. 3. The blue line indicates the user's motion trajectory and the red arrows attached to the blue line indicate the gaze direction at the corresponding position. Not surprisingly, we can find that **users' gaze often follows the activity of the object inside the video**. Specifically, they can be divided into two categories. On the one hand, **for volumetric scenes with relatively large movement, users' gaze tends to precede users' movement by a short period of time**. This phenomenon can be observed from Fig. 3 (a) and Fig. 3 (b), where the trolley and dustpan follow a regular movement. Then users' gaze can be focused on these objects and appear to have a similar movement feature. On the other hand, **for volumetric scenes with small movements, the gaze may move back and forth with irregular movement, but it generally still focuses on the target object**. Fig. 3 (c) and Fig. 3 (d) verify this observation that the endpoints of the gaze arrows mostly locate at the target objects.

Fig. 4 reaffirms this observation. The left figure shows the original aerial view heatmap of the movement trajectory, and the light part of the right figure indicates the point where the movement trajectory coincides with the gaze ray. It demonstrates a strong correlation that a large portion of the movement trajectory and the gaze ray indeed have interaction.

## 4 ANALYSIS ON USER BEHAVIOR

In this section, we conduct a comprehensive analysis of user behaviors based on the dataset, aiming to reveal the implicit correlations between various observed features, and further provide insight for future user behavior prediction. We mainly focus on user attention and movement features.

### 4.1 Volumetric ROI Calculation

Users' region of interest (ROI) is the most important feature when viewing volumetric videos. However, different from 2D or 360-degree videos where ROI can be directly obtained, ROI calculation in volumetric video is not so intuitive given its 3D nature. On one hand, there can be multiple objects alongside a user's eyesight and it is hard to uniquely determine the interested object. On the other hand, users are moving most of the time and the viewing angles are constantly changing. Thus, we define the *volumetric ROI level*,

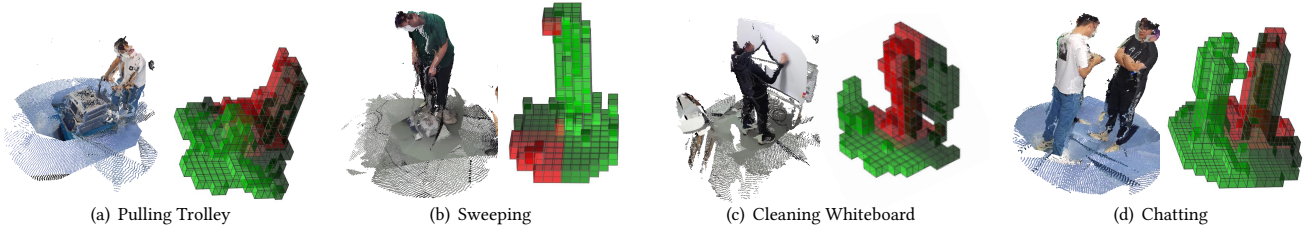

(a) Pulling Trolley  (b) Sweeping  (c) Cleaning Whiteboard  (d) Chatting

Figure 5: The volumetric ROI level together with 4 representative scenes. Here the light green color indicates a higher ROI level and the dark red color indicates a lower ROI level.

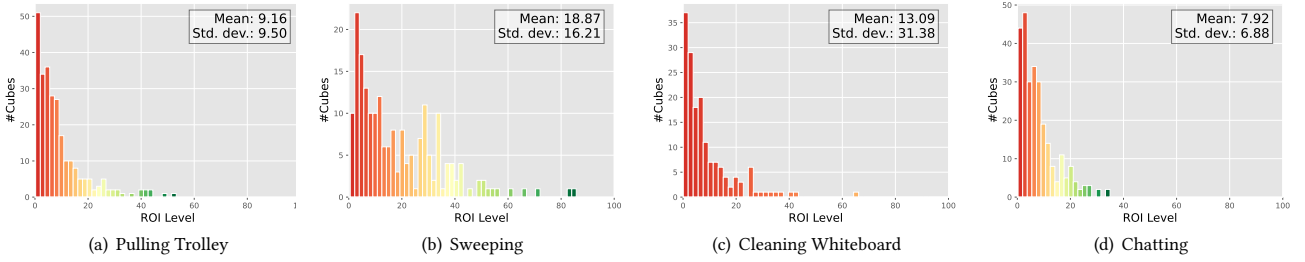

(a) Pulling Trolley  (b) Sweeping  (c) Cleaning Whiteboard  (d) Chatting

Figure 6: The distribution of the volumetric ROI level of four scenes. X-axis indicates the ROI levels, and Y-axis indicates the number of cubes with the corresponding ROI.

as a quantitative indicator, to represent how much attention a user pays to a region.

Calculating the volumetric ROI level includes the following steps:

▷ **Scene segmentation.** We first divide the whole volumetric scene into small blocks, where each block is a cube after slicing the space from x, y, and z dimensions. Since most of the cubes do not contain any points or only contain very few points, we set a threshold to filter out those near-empty cubes and only preserve those representing practical objects. Note that users' sensitivity to the point cloud density decreases with the increase of observing distance [8], we also vary such threshold accordingly.

▷ **Gaze frustum calculation.** By exploiting the pre-processed headset trajectory and gaze data, we are able to calculate the viewing directions of the user at every position. Normally, people's effective viewing angle is about 30° [27, 29, 32], we therefore define a virtual viewing frustum with an angle of 30°. And objects within this frustum will be viewed by the user.

▷ **Intersection calculation.** The ROI level of one cube can be calculated as how frequently this cube is covered by the gaze frustum of the user. In practice, we calculate the direction vector formed by the coordinates of the headset and the center of the cubes and then compare the angle between the direction vector and the gaze direction vector obtained from previous processing. The cube is counted once every time the angle is less or equal to 30°. By going through all of the effective cubes, we obtain the total counts for the whole volumetric video.

▷ **Volumetric ROI level calculation.** Inspired by the ROI mechanism used in 360 videos [5], we propose to calculate the volumetric ROI level $F_a$ of a cube according to the density weight, the appearance frequency, and the distance between the user's eyes and the cube. The calculation formula is given as:

$$F_a = \frac{\rho_c * f_g}{D_c} \quad (3)$$

$$f_g = \frac{\sum_{i=1}^{N} N_g(i)}{N_{sample}} \quad (4)$$

where $\rho_c$ is the point cloud density of the cube, $f_g$ is the frequency of each cube falling into the viewing frustum, $D_c$ is the distance between the headset and the cube center, $N_g(i)$ is the total counts of the current cube, and $N_{sample}$ is the total number of user behavior samples.

### 4.2 Analysis on User Attention

We next analyze users' attention (which can be directly reflected by the ROI level) when they are viewing different volumetric videos. Fig. 5 visually shows the different ROI levels for different volumetric video scenes. Here we randomly select 5 viewers and illustrate their average ROI level. We can find that users' attention is highly correlated with the volumetric content, and is **particularly on the actors and the objects they are manipulating**. For example, in the 'Chatting' scene, most attention is focused on the right person. In the 'Sweeping' scene, the dustpan instead attracts even more attention than the person.

Another interesting finding lies in the personalized preference, i.e., **users may pay higher attention to their preferred object or person**. Like in Fig. 5(d), the right person obviously has a higher ROI level than the left person, which is largely due to the user's personalized preference.

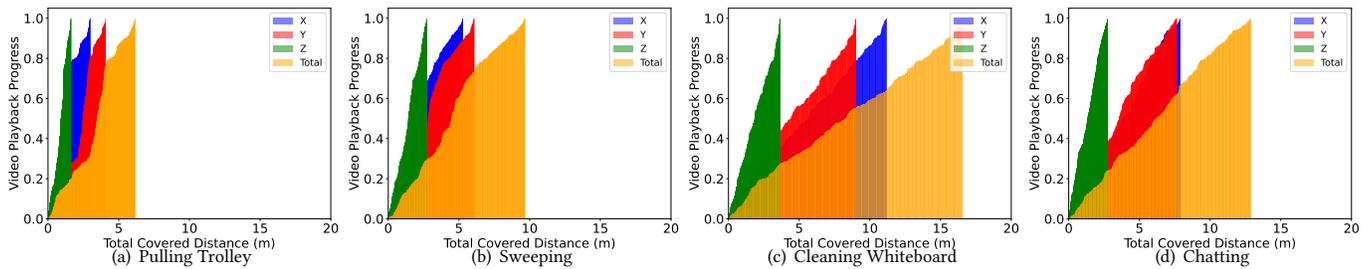

Figure 7: Movement Distance of X, Y, Z axes

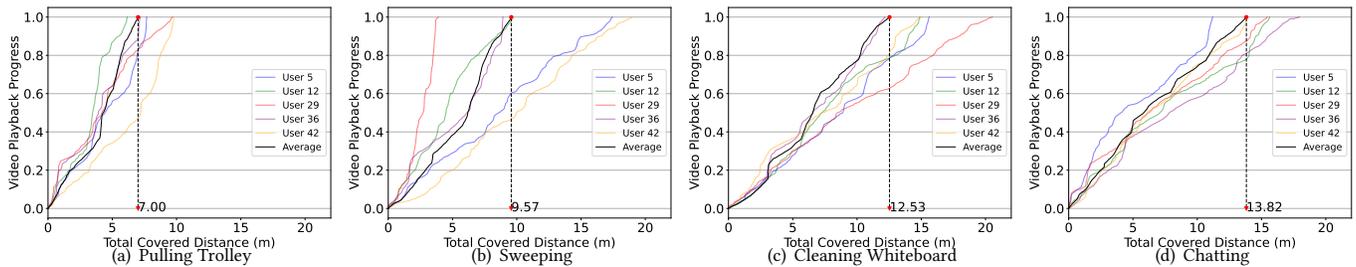

Figure 8: The total movement distance of different scenes. Here the colored line and the black line indicate each user's total movement distance and the average respectively along the video playback progress, the dotted line with a number represents the average value.

We next consider the distribution of user attention in different volumetric scenes. Fig. 6 plots the distribution of cubes with different volumetric ROI levels together with the mean value (Mean) and the standard deviation (Std. dev.). Note that we already remove those rarely-watched cubes. Comparing the different volumetric scenes, we can obtain several interesting findings: 1) **The ROI dispersion level of different volumetric videos is quite diverse, depending on the scene content**. For example, the ROIs of the 'Sweeping' scene concentrate with the range from 0 to 60, while the ROIs of the 'Chatting' scene mainly spread between 0 to 15. This means that users are more focused when watching the former more 'dynamic' video while they are more distracted when watching the latter more 'static' one. And it further reaffirms that people's attention is more easily captured by moving objects. 2) **Only a small portion of cubes have relatively high ROI levels.** This is because a volumetric scene can have a lot of effective cubes, while only a small portion of them, especially those representing the target actors or objects, will gain enough attention.

### 4.3 Analysis on User Movement

We conduct a more in-depth analysis of user movement to examine the correlations between movement behavior and video content. We first define the movement mode. Taking the user's headset as the origin, moving along the lateral direction of the body is indicated as the x-axis, along the vertical direction of the body is indicated as the y-axis, and the z-axis represents the up-down movement.

Fig. 7 shows the average moving distance along video playback progress in the three directions as well as the total distance of the 4 volumetric scenes. Naturally, moving laterally means that the user prefers to observe from different angles while moving vertically means that users would like to follow the moving objects. From this figure, we can find that the vertical distance is clearly larger than the lateral distance in the 'Pulling Trolley' and 'Sweeping' scenes, and vice versa for the rest two scenes. This observation matches exactly with our previous finding that **people tend to follow the moving object while observing the static object from various angles**.

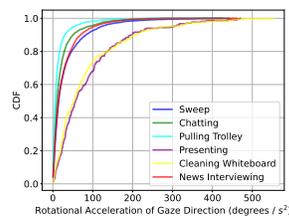
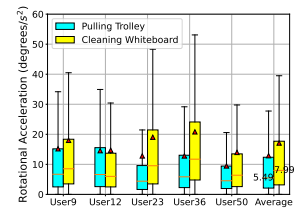

Figure 9: Rotational Acceleration in Different Scenes

Figure 10: Rotational Acceleration of 'Pulling Trolley' and 'Cleaning Whiteboard'

We also investigate the movement features from the perspective of different users. Fig. 8 shows the cumulated moving distances of five randomly selected users. We find that the first two dynamic-scene videos have an average moving distance of 7.0m and 9.57m, respectively, while the rest two static-scene videos reach an average moving distance of 12.53m and 13.82m. Thus we can verify that **users tend to perform more spatial movements in static scenes compared to dynamic scenes to explore more areas in volumetric scenes**.

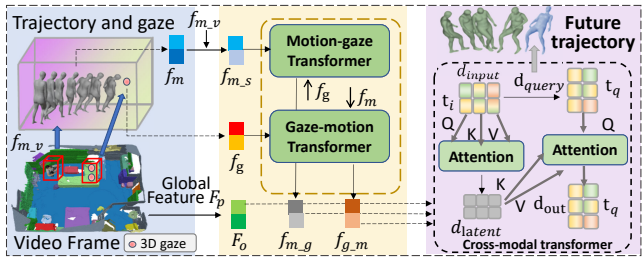

Figure 11: Transformer-based viewport prediction model.

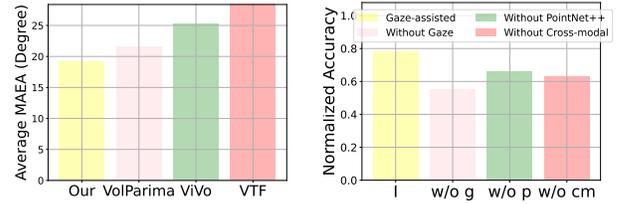

Figure 12: Average MAEA for viewport prediction

Figure 13: Average Accuracy for each prediction part

In Fig. 9, we show the average data to observe the difference in rotational acceleration for different scenes. From the figure, we find that in more static scenes, users change their orientation more frequently at a faster rate. According to the CDF plot, for even more than 30% sampling points the head moving speeds exceed 100 degree/$s^2$ in the 'cleaning whiteboard' scene. In contrast, for the relatively dynamic scene 'Pulling Trolley', there are about 80% sampling points with moving speed less than 20 degree/$s^2$. Diversity across different users is shown in Fig. ?? and Fig. ??, which depicts the specific rotational acceleration speed of 5 randomly selected users and their average for the scenes of 'Pulling Trolley' and 'Cleaning Whiteboard'. Observation from these figures reaffirms the finding that user movement in static scenes is usually faster than in dynamic scenes.

## 5 GAZE-ASSISTED VIEWPORT PREDICTION FOR VOLUMETRIC VIDEO STREAMING

In this section, we give a case of the dataset application in volumetric video streaming. By fusing the correlated features between video content and gaze information, we are able to improve the accuracy of viewport prediction, further benefiting VV streaming.

### 5.1 Background and Motivation

Viewport adaptive video streaming together with tile-based partition strategy [8, 15] has been widely explored in traditional 2D and recently 360-degree videos. By reducing the bitrate of the video content outside users' viewport, the whole transmitted video size can be saved and thus relieving the network bandwidth pressure. This idea is intuitive to move to the 3D scenario if the scene is partitioned into small cubes for cube-based streaming. However, though it applies well in 2D videos, a critical challenge arises when it comes to 3D volumetric videos. The major difficulty lies in the flexible six DoF spatial feature, where the significant uncertainty in spatial position and viewing angle makes the viewport prediction error easy to accumulate.

Several pioneer works have made attempts for six DoF viewport prediction [6, 15, 25]. For example, ViVo [8] and Vues [19] employ linear regression (LR) and multilayer perceptron (MLP) to predict the viewport, and have also explored the use of advanced deep learning models such as LSTM for prediction. Extending from Parima [1], VolParima [18] utilizes 3D object detection and tracking techniques to achieve improved accuracy in viewport prediction. However, these works either consider each DoF separately or mainly focus on the video content, which cannot fully capture the implicit features in volumetric videos to yield accurate viewport prediction towards various volumetric scenes. Motivated by our previous observations and findings, **we realize that the features in user movement, gaze direction, and video content are tightly correlated so that the multi-modal information, as well as their mutual impacts, should be combined together for consideration.**

### 5.2 Design

We extract the multimodal features and present an architecture with a bidirectional fusion model that facilitates the communication of different features in Fig. 11. This is a paradigm for accurate viewport predictions based on video content, interaction, and intention. Followed by a variety of cross-modal transformers to transcend information from multi-modality.

**Cross-modal transformer.** The cross-modal transformer [10] is used to capture the interplay of several elements and to establish communications among the multi-modal information.

Instead of extracting the multi-modal features independently[14], we propose a pipeline to overall integrate the history viewport feature, 3D gaze feature, and video features, which enhances the in-between feature communication to mutually decrease their future uncertainties on interaction and intention.

**Video feature extraction.** To learn the constraints (e.g. Surface and topology of furniture) from the 3D video and retrain the network for attention on locally interacted structures, we apply Point-Net ++ [24], to extract both global (the video content) and local video features (interacted region). We derive the per-point feature and global descriptor of video as $F_P, F_o$.

**Gaze feature extraction.** The gaze point feature $f_g$ is retrieved from the per-point video feature map $F_P$ into $F_{P|g}$. Consequently, the interacted gaze feature with corresponding video information provides indications to infer the intention.

**Viewport feature extraction.** We use a linear layer to extract the viewport feature embedding $f_m$ from multidimensional viewport trajectories input. The viewport is well-aligned with the video content. To endow the feature awareness of the 3D video content, we further query the video features with the viewport features. These interacted video features are then supplied to PointNet++ to get the contextual video feature $f_{m-v}$ of the current viewport.

In lieu of directly concatenating the features, which would bring modalities features redundancy and impair the prediction accuracy [16], we propose a model by deploying a cross-modal transformer [22] to fuse the gaze, viewport, and video features.

**Feature fusion.** As an intermediary element, the viewport features strive to be cognizant of the 3D video features and the subject's intention inferred from the gaze features. First, we utilize the video feature $f_{m\_v}$ acquired from the 3D environment as the query to update the viewport feature $f_m$ in the viewport-video transformer. Then, the output viewport embedding $f_{m\_s}$ is expected to be aware of the 3D video, which results in the final viewport embedding $f_{m\_g}$. Inspired by [35], we handle the gaze embedding in a bidirectional manner, i.e., the viewport embedding $f_m$ is also utilized as the query to update the gaze features into $f_{g\_m}$. The bidirectionally fused multi-modal features are then assembled into holistic temporal input representations to perform human viewport prediction. As shown in Fig. 11, the updated gaze feature $f_{g\_m}$, viewport feature $f_{m\_g}$ and the global video feature $F_O$ are used to predict the future viewport trajectories from $t$ to $T$ by:

$$V_{T:T+t} = \Re\left(h_{\text{pos}}, \text{concat}\left(f_{g\_m}, f_{m\_g}, F_O\right)_{T-n:T-1}\right) \quad (5)$$

where concat denotes operator of concatenation, and $h_{\text{pos}}$ is the latent vector containing temporal positional encodings for the output[24]. We evaluate our gaze-assisted viewport prediction against representative VV system and methods `ViVo`, `VolParima` and transformer-based `Vanilla-TF (VTF)` [31] using the Average Mean Absolute Error Angle (MAEA) as a metric. We also do an ablation study to compare the effect of each part.

As depicted in Figure 12, our proposed model is capable of reducing MAED by 13.3%, 19.8%, and 34.5% in comparison with `VolParima`, `ViVo`, and `VTF`, respectively. Furthermore, we conducted experiments to evaluate the accuracy of our gaze-assisted model and performed an ablation study comprising three variations: without gaze (w/o g), without PointNet ++ (w/o p), and without a cross-modal transformer (w/o cm). The results indicate that each component has a positive contribution to the overall performance. Our model, which effectively integrates and utilizes video content and gaze information, is demonstrated to produce more accurate predictions than the previous methods.

## 6 OTHER APPLICATIONS

In addition to our proposed viewport prediction systems, we provide several potential application cases that could be derived from our dataset.

### 6.1 User Identification for VV

User identification is a crucial task in 360-degree video, yet it poses a new challenge for volumetric video. Such a technique has the potential to improve user experience or enhance privacy.

For headset-movement-based identification, Li et al. [17] achieved an identification accuracy of 95.57% while participants nodded when listening to music. Gaze data could also be used for identification, Sluganovic et al. [28] proposed gaze-based authentication using a gaze-tracking device, their system his system achieves an error rate of 6.3% at an authentication time of 5 seconds.

Given that our dataset on VV user behavior encompasses a wider range of attributes, an identification method utilizing both headset and gaze data could be developed to enhance accuracy.

### 6.2 Personalized Content Delivery

Many works have been conducted for content recommendation traditional in 2D video [3, 4] and 360-degree video [30], but for volumetric video, such field is still undefined. By analyzing the behavior of the users, developers can gain insights into users' preferences and adapt personalized content to better suit their needs.

**Content Recommendation.** Based on the historical movement pattern of the users and the viewing history, developers can build user portraits for each user, in order to deliver new VV contents that are more likely to be of interest to the user.

**Adaptive Content** Using the insights gained from analyzing the users' behavior, developers are able to dynamically adjust the VV experience in real time. For example, the lighting could be adjusted when the users tend to change their viewport frequently to minimize motion sickness.

### 6.3 Healthcare

VV user behavior analysis has the potential to play a role in psychoanalysis, particularly in the area of virtual reality therapy.

By analyzing changes in users' behavior and movement patterns before and after virtual reality therapy, therapists can evaluate the effectiveness of the treatment. For example, if a patient with a fear of heights spends more time looking down from a virtual high-rise building after therapy than before, this suggests that the treatment has been effective.

## 7 CONCLUSION

In this paper, we focused on understanding user behavior patterns when watching volumetric videos. We released the first large-scale volumetric video user behavior dataset, including movement information, headset direction, user gesture, and user gaze information. This dataset involved data from 50 users with strong diversity and covered multiple representative volumetric scenes. We then conducted a comprehensive analysis aiming to reveal the behavior features. We defined the volumetric ROI level calculation mechanism in this context and focused on the feature analysis on user attention and user movement. Some interesting findings were therefore derived. Further, based on our analysis and observation, we designed a transformer-based volumetric video viewport prediction model, which fused all the correlated features and outperformed the state-of-the-art baseline solutions.

## ACKNOWLEDGMENTS

The work was supported in part by the Basic Research Project No. HZQB-KCZYZ-2021067 of Hetao Shenzhen-HK S&T Cooperation Zone, by NSFC (Grant No. 62293482 and No. 62102342), the Guangdong Basic and Applied Basic Research Foundation (Grant No. 2023A1515012668), the Shenzhen Science and Technology Program (Grant No. RCBS20221008093120047), the Shenzhen Outstanding Talents Training Fund 202002, the Guangdong Research Projects No. 2017ZT07X152 and No. 2019CX01X104, the Guangdong Provincial Key Laboratory of Future Networks of Intelligence (Grant No. 2022B1212010001), the Shenzhen Key Laboratory of Big Data and Artificial Intelligence (Grant No. ZDSYS201707251409055).